\documentclass[aps,twocolumn,prl,superscriptaddress]{revtex4-1}

\usepackage{csquotes}
\usepackage{amsmath}
\usepackage{amssymb}
\usepackage{color}
\usepackage{hyperref}
\usepackage{bm}
\usepackage[normalem]{ulem}
\usepackage{hhline}

\usepackage{graphicx}

\DeclareMathAlphabet{\mathitbf}{OML}{cmm}{b}{it}

\renewcommand{\th}{^{\mbox{\tiny th}}}

\newcommand{\dbar}{{\,\mathchar'26\mkern-12mu d}}

\begin{document}

\title{Universal non-phononic density of states in 2D, 3D and 4D glasses}
\author{Geert Kapteijns}
\affiliation{Institute for Theoretical Physics, University of Amsterdam, Science Park 904, 1098 XH Amsterdam, The Netherlands}
\author{Eran Bouchbinder}
\affiliation{Chemical and Biological Physics Department, Weizmann Institute of Science, Rehovot 7610001, Israel}
\author{Edan Lerner}
\affiliation{Institute for Theoretical Physics, University of Amsterdam, Science Park 904, 1098 XH Amsterdam, The Netherlands}

% \title{Universal form of the non-phononic density of states \\ in 2D, 3D and 4D structural glasses}
% \author{Geert Kapteijns${}^{1}$, Eran Bouchbinder${}^{2}$, and Edan Lerner${}^{1}$}
% \affiliation{${}^1$Institute for Theoretical Physics, University of Amsterdam, Science Park 904, 1098 XH Amsterdam, The Netherlands \\ ${}^2$Chemical and Biological Physics Department, Weizmann Institute of Science, Rehovot 7610001, Israel}

\begin{abstract}
It is now well established that structural glasses possess disorder- and frustration-induced soft quasilocalized excitations, which play key roles in various glassy phenomena. Recent work has established that in model glass-formers in three dimensions, these non-phononic soft excitations may assume the form of quasilocalized, harmonic vibrational modes whose frequency follows a universal density of states $D(\omega)\!\sim\!\omega^4$, independently of microscopic details, and for a broad range of glass preparation protocols. Here we further establish the universality of the non-phononic density of vibrational modes by direct measurements in model structural glasses in two dimensions and four dimensions. We also investigate their degree of localization, which is generally weaker in lower spatial dimensions, giving rise to a pronounced system-size dependence of the non-phononic density of states in two dimensions, but not in higher dimensions. Finally, we identify a fundamental glassy frequency scale $\omega_c$ above which the universal $\omega^4$ law breaks down.
\end{abstract}

\maketitle

\emph{Introduction.---} Understanding the statistical and structural properties of low-frequency excitations in disordered solids is a long-lasting challenge in condensed matter physics \cite{soft_potential_model_1991,Schober_prb_1992,Gurevich2003,Schober_Laird_numerics_PRB,ohern2003,barrat_3d, Schirmacher_prl_2007, Monaco_prl_2011, sokolov_boson_peak_scale, mw_EM_epl, eric_boson_peak_emt, silvio, Zamponi, modes_prl, phonon_widths}. It is generally accepted that in addition to the well-understood long-wavelength phonons that dwell at low frequencies, other soft excitations that stem from different aspects of the microstructure and/or the disorder may emerge. These excitations have been argued to play an important role in determining transport \cite{Schober_prb_1992,Schirmacher_prl_2007,eric_boson_peak_emt}, static \cite{exist,Biroli2016} and thermodynamic \cite{soft_potential_model_1991,eric_boson_peak_emt} properties of disordered solids, as well as to control relaxation patterns in equilibrium supercooled liquids \cite{wyart_brito_2007,widmer2008irreversible,wyart_brito_2009,harrowell_2009}, and plastic flow rates in externally-deformed glasses \cite{manning2011,lte_pnas}.

One of the origins of the non-phononic low-frequency vibrational modes is related to the degree of connectedness of the underlying network of (strong) interactions formed between the constituent particles of a disordered solid. Upon decreasing the connectedness of the aformentioned network, which can occur e.g.~by carefully decompressing packings of soft repulsive spheres \cite{ohern2003}, the rigidity of a glass becomes gradually compromised, a phenomenon known as `unjamming'~\cite{liu_review,van_hecke_review}, and is accompanied by the appearance of low-frequency non-phononic vibrational modes whose characteristic frequency vanishes when rigidity is completely lost. This emergence of soft vibrational modes, as well as many other aspects of the unjamming phenomenology, %phenomenology associated with unjamming,
are well-captured by variational arguments \cite{eric_hard_spheres_emt,new_variational_argument_epl_2016} and mean-field theories \cite{mw_EM_epl,eric_boson_peak_emt,Zamponi, silvio}.

%%%%%%%%%%%%%%%%%%%%%%%%%%%%%%%%%%%%%%%%%%%%%%%%%%%%%%%
\begin{figure}[!ht]
\centering
\includegraphics[width = 0.5\textwidth]{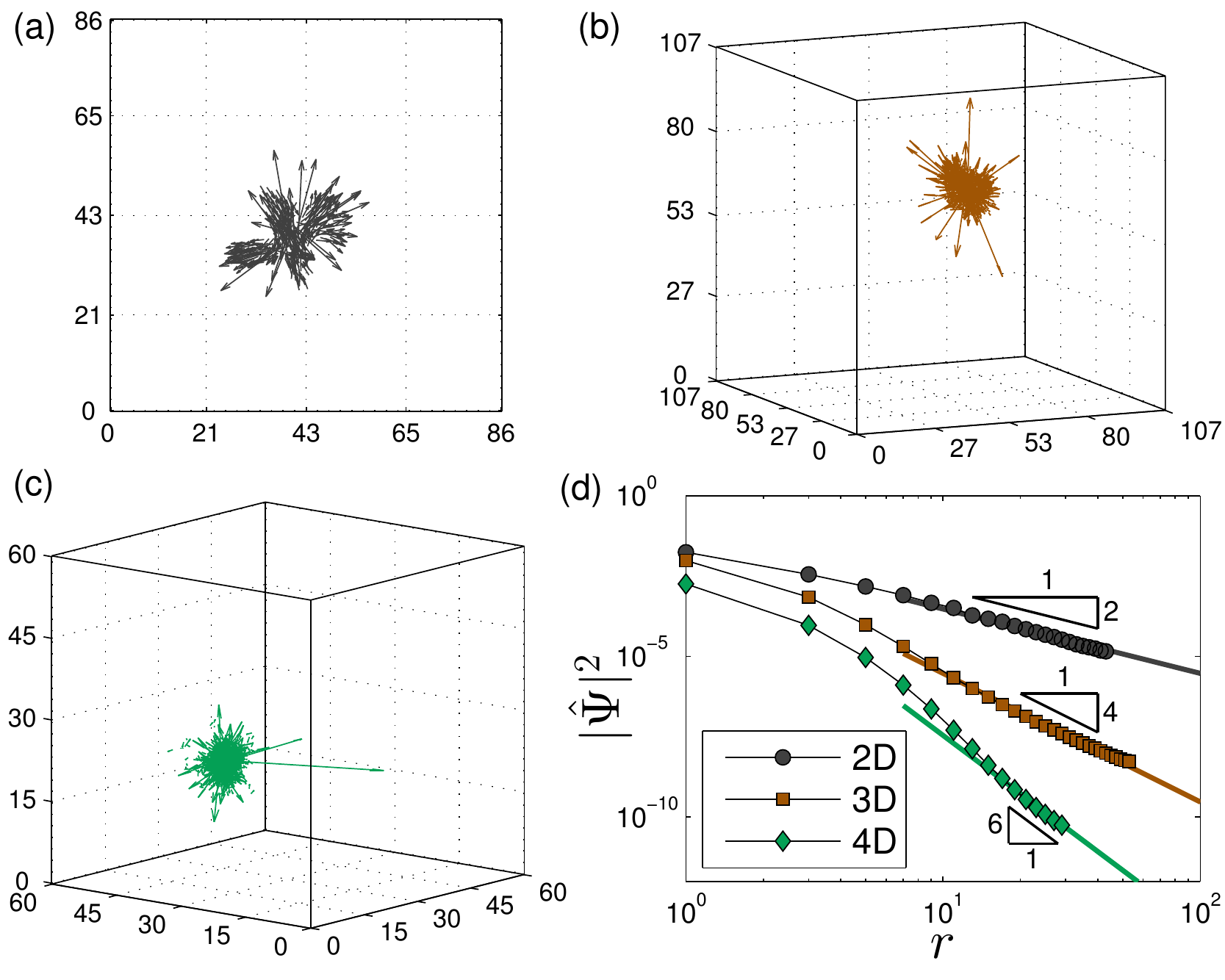}
\caption{\footnotesize Visualizations of soft quasilocalized vibrational modes in (a) 2D (b) 3D and (c) a 3D projection of a 4D mode. We plot the largest 3\%, 0.1\% and 0.01\% components, that amount to approximately 70\%, 92\% and 87\% of the displayed modes' weights in 2D, 3D and 4D, respectively. The modes displayed have frequencies of roughly half the respective longest wavelength transverse phonon frequency or less, and were calculated in a generic model glass of purely repulsive spheres interacting via a $r^{-10}$ pairwise potential, see text for details.  (d) Symbols represent the spatial decay of the squared magnitude (see \cite{plastic_modes_prerc} for details) of the modes shown in panels (a)-(c), as a function of the distance $r$ away from the modes' respective cores.} %\sout{The continuous lines follow $r^{-2(\dbar-1)}$ in $\dbar$ spatial dimensions, see text for further discussion.} }
\label{introduction_fig}
\end{figure}
%%%%%%%%%%%%%%%%%%%%%%%%%%%%%%%%%%%%%%%%%%%%%%%%%%%%%%

In a broad class of glass-forming models, referred to in what follows as `generic' glass-forming models \footnote{One prominent example of a generic glass-forming model is the Kob-Andersen Binary Lennard-Jones model~\cite{kablj}, which is arguably the most widespread model in contemporary numerical studies of glass physics}, the unjamming phenomenology is often irrelevant, as these systems dwell far away from the critical unjamming point \cite{stefanz_pre}. Notwithstanding, generic model glasses still feature soft non-phononic excitations; examples of such excitations are presented in Fig.~\ref{introduction_fig}. Recent investigations of generic glass forming models in three dimensions (3D) have established the following general observations:

\begin{enumerate}

\item{The softest non-phononic excitations in generic glass models are quasilocalized: they feature a spatial structure consisting of a disordered core of size $\xi$ (on the order of 10 particle diameters), decorated by a power-law decay $\sim\! r^{-2}$ at distances $r$ away from the core \cite{plastic_modes_prerc, modes_prl, SciPost2016}, as demonstrated in Fig.~\ref{introduction_fig}.} %The participation ratio (a conventional quantifier of the degree of localization of an excitation, see definition below) of soft quasilocalized excitations scales as $N^{-1}$ in systems of $N$ particles \cite{modes_prl, SciPost2016}.}

\item{Due to strong hybridizations with long-wavelength phonons (elastic waves), soft quasilocalized excitations may only assume the form of harmonic vibrational modes (i.e.~eigenfunctions of the dynamical matrix, see definition below) %  in certain system-size-dependent frequency ranges. This may happen in particular
when (i) systems are made small enough to suppress the occurrence of phonons at very low frequencies \cite{modes_prl, SciPost2016, inst_note}, but still larger in linear size than the core size $\xi$, or (ii) system sizes are employed for which a coexistence frequency window of non-phononic quasilocalized vibrational modes and phonons opens. Within this coexistence frequency window, which has been shown in \cite{phonon_widths} to vanish in the thermodynamic limit, quasilocalized excitations can assume the form of harmonic vibrations if their frequencies fall in the gaps between neighboring phonon bands, as observed in \cite{ikeda_pnas,Ikeda_PRE_2018}. The proximity of phonons with similar frequencies to quasilocalized vibrational modes' frequencies can alter the latter's far-field structure, which may feature a phononic background \cite{Schober_jop_2004,Schober_Ruocco_2004,ikeda_pnas} instead of a power-law decay away from their cores. We note that the effect of quasilocalized modes on various phenomena in glasses is not expected to depend on their observability as harmonic vibrational modes, as discussed e.g.~in \cite{soft_potential_model_1991,lte_pnas}.}

\item{The existence of soft quasilocalized excitations depends crucially on the presence of frustration-induced internal stresses \cite{ikeda_pnas, inst_note}, which are a generic feature of structural glasses \cite{shlomo}.}

\item{When soft quasilocalized excitations are realized as harmonic vibrational modes, their frequencies follow a universal non-phononic distribution $D(\omega)\!\sim\!\omega^4$ \cite{modes_prl}, which persists over a broad range of glass preparation protocols \cite{protocol_prerc, inst_note, cge_paper}. Several theoretical frameworks predicted this universal distribution \cite{soft_potential_model_1991,Schober_prb_1992,Gurevich2003}.}

\end{enumerate}

While important progress in elucidating the degree of universality of the non-phononic vibrational density of states (vDOS) has been recently made \cite{modes_prl, SciPost2016, protocol_prerc, inst_note, phonon_widths, marco2015prl}, several fundamental questions remain open: (i) is the $\omega^4$ law universal across spatial dimensions? (ii) are there clear signatures of dimension-dependent finite-size effects on the non-phononic vDOS, and, if so, what is their origin? (iii) does a frequency scale exist above which the universal $\omega^4$ law breaks down? %If so, does it represent a fundamental glassy energy scale?

In this Letter we address these pressing questions; we show that the form of the non-phononic vDOS $D(\omega)\!\sim\!\omega^4$, which has been previously observed only in 3D models, persists in two dimensional (2D) and four dimensional (4D) model glasses as well, firmly establishing its universality. We also directly demonstrate the quasilocalized nature of these non-phononic modes. These results cast doubt on the claims that the effective medium theory~\cite{eric_boson_peak_emt} and the infinite-dimensional mean-field theory~\cite{silvio}, which predict a different scaling, are relevant to realistic dimensions away from unjamming~\cite{non_debye_prl_2016}. We further demonstrate that the degree of localization of QLVM is weaker in lower dimensions, giving rise to a pronounced system-size dependence of the non-phononic vDOS in 2D. Finally, by exploiting the scaling of phonon frequencies and the stronger localization of QLVM in 4D, our analysis reveals an emergent cutoff frequency $\omega_c$ above which the $\omega^4$ law breaks down. We speculate about the implications of $\omega_c$ and mention future research directions.

\emph{Model glass and methods.---} We employ a well-studied glass-forming model that consists of a 50:50 mixture of `large' and `small' particles of mass $m$ in 2D, 3D and 4D. $N$ particles are enclosed in a $\dbar$-dimensional hypercube of edge length $L$ with periodic boundary conditions, and interact via a radially-symmetric purely repulsive pairwise energy that varies with the distance $r$ between particles proportionally to $\epsilon (\lambda/r)^{10}$, with $\lambda$ and $\epsilon$ denoting our microscopic units of length and energy, respectively.  A complete description of the model can be found e.g.~in \cite{cge_paper}. In what follows all observables should be understood as expressed in terms of the appropriate microscopic units. We choose the densities $\rho\!\equiv\! m N / L^\dbar$ to be 0.86, 0.82, and 0.80 in 2D, 3D and 4D, respectively, such that the first peak of the radial distribution function $g(r)$ is approximately aligned across dimensions. Glassy samples in all dimensions were created by equilibrating the fluid phase at $T\!=\!1.0$ and then cooling systems deep into the glassy phase (down to $T\!=\!0.1$) at a constant rate of $\dot{T}\!=\!10^{-3}$. Residual heat was then removed by performing an energy minimization. The number of glassy samples generated for each ensemble is detailed in Table~\ref{ensemable_sizes_table}. Vibrational modes were calculated by a partial diagonalization of the dynamical matrix ${\cal M}\!\equiv\!\frac{\partial^2 U}{\partial \vec{x} \partial \vec{x}}$, where $\vec{x}$ denotes particles' coordinates. The Debye frequencies (defined e.g.~in \cite{kittel1996introduction}) are found to be $\omega_D\!\approx$ 20.2, 17.3, and 23.3 in 2D, 3D and 4D, respectively.
\begin{center}
\begin{table}
\resizebox{0.49\textwidth}{!}{
\begin{tabular}{ c|cccc|ccc|ccc }
 \hline
 & \multicolumn{4}{c|}{2D} & \multicolumn{3}{c|}{3D} & \multicolumn{3}{c}{4D} \\
 \hline
 $N$ & 196  &  400 & 786 & 1600 & 1K & 2K & 4K & 5K & 10K & 20K  \\
 \hline
 ensemble size & 1M & 490K & 250K & 122.5K & 50K & 50K & 100K & 4K & 2K & 1K \\
 \hline
 \end{tabular}
 }
 \caption{\footnotesize System and ensemble sizes of glassy samples generated.}
 \label{ensemable_sizes_table}
\end{table}
\end{center}

%%%%%%%%%%%%%%%%%%%%%%%%%%%%%%%%%%%%%%%%%%%%%%%%%%%%%%%
\begin{figure*}[!ht]
\centering
\includegraphics[width = 0.9\textwidth]{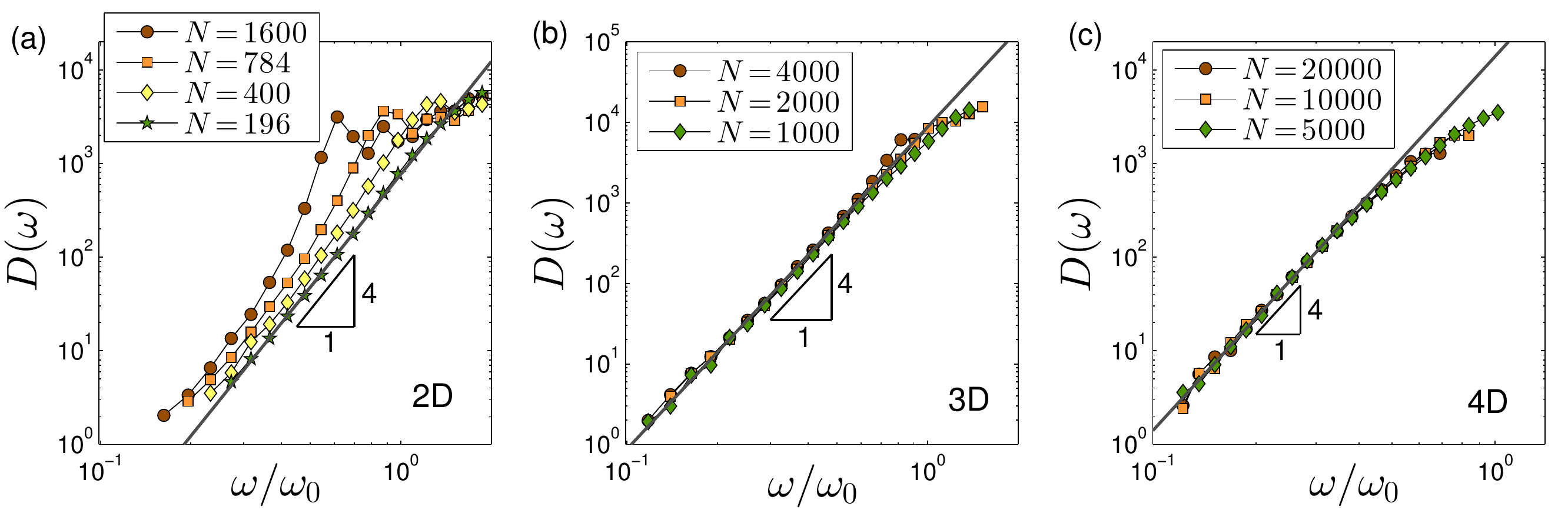}
\caption{\footnotesize Sample-to-sample average of the density of vibrational modes $D(\omega)$, in (a) 2D, (b) 3D, and (c) 4D glasses, for a variety of system sizes as indicated by the legends. We find $D(\omega)\!\sim\!\omega^4$ in all spatial dimensions. The intrusion of phonons into the $\omega^4$ scaling regime upon increasing system size is clearly observed in our 2D data. Noticeably, the prefactor of the $\omega^4$ scaling appears to be system-size independent in 3D and 4D, but not so in 2D, where it grows with increasing system size, see text for discussion. The frequency axes are rescaled by $\omega_0\!=1.0, 1.6, 2.8$ for 2D, 3D and 4D, respectively, for visualization purposes. The vertical axes are rescaled such that $D(\omega)\!\sim\!1$ for the lowest frequencies measured.}
\label{all_dimensions_dos}
\end{figure*}
%%%%%%%%%%%%%%%%%%%%%%%%%%%%%%%%%%%%%%%%%%%%%%%%%%%%%%

\emph{Results.---} Figure~\ref{all_dimensions_dos} displays our main result; in panels (a),(b), and (c) we display the ensemble-average vDOS $D(\omega)$ measured in our 2D, 3D, and 4D glassy samples, respectively. At low frequencies we find
\begin{equation}
D(\omega)\!\sim\! \omega_g^{-5}\omega^4\,,
\end{equation}
in all spatial dimensions studied, firmly establishing the universality and fundamentality of this emergent law. We further find that $\omega_g\!=\!\omega_g(N;\dbar)$ is a dimension- and system-size-dependent prefactor \footnote{$\omega_g$ also depends on the preparation protocol of the glass, as discussed at length in~\cite{cge_paper}}. Noticeably, the dependence of $\omega_g$ on $N$ is much stronger in 2D compared to 3D and 4D.

Where does the pronounced system-size dependence of the vDOS in 2D emanate from? In Fig.~\ref{introduction_fig} we show examples of vibrational modes that populate the $\omega^4$ tails of the vDOS across all investigated spatial dimensions. Consistent with previous observations in 2D and 3D~\cite{plastic_modes_prerc, modes_prl, SciPost2016}, we find that these modes feature a disordered core of linear size on the order of 10 particle diameters, decorated by a power-law decay $\sim\! r^{-(\dbar-1)}$ at distances $r$ away from the core. These power-law spatial decays lead us to expect that the characteristic frequency scales of QLVM should feature an $N$ and $\dbar$ dependence, that would in turn translate to $N$- and $\dbar$-dependent prefactors $\omega_g^{-5}$ of the $\omega^4$ vDOS.

To estimate $\omega_g(N;\dbar)$ we consider similar objects --- the linear displacement response fields to local force dipoles~\cite{cge_paper, breakdown}; such responses feature similar spatial structures to those of QLVM \cite{cge_paper}, and in particular their far fields also scale as $r^{-(\dbar-1)}$ at distances $r$ away from the force dipole, just as seen for QLVM in Fig.~\ref{introduction_fig}. The stiffness (frequency squared) associated with these fields was derived in~\cite{cge_paper} using the elastic Green's function in $\dbar$ dimensions; the result reads
\begin{equation}
\omega_g(N;d) \sim \left\{\begin{matrix}(\log N)^{-1/2}\,,&\dbar=2\\ \omega_{g,\infty} + A\,N^{-(\dbar-2)/\dbar}\,,& \dbar > 2\end{matrix} \right. \,.
\end{equation}
This results predicts that the prefactor of the non-phononic vDOS grows without bound as $\omega_g^{-5}\!\sim\!(\log N)^{5/2}$ in 2D, but converges to a constant for large $N$ in $\dbar\!>\!2$, in qualitative agreement with our observations.

We next turn to a statistical analysis of the degree of localization of QLVM across dimensions and system sizes. To this aim we consider the participation ratio $e$ of a general field $\vec{z}$, defined as
\begin{equation}
e \equiv \frac{\big(\sum_i \vec{z}_i\cdot\vec{z}_i\big)^2}{N\sum_i (\vec{z}_i\cdot\vec{z}_i)^2}\,,
\end{equation}
where $\vec{z}_i$ denotes the $\dbar$-dimensional Cartesian components of $\vec{z}$ pertaining to the $i\th$ particle. The participation ratio is a general indicator of the degree of localization of a field: extended fields feature participation ratios of order unity, whereas $e\!\sim\!N^{-1}$ for localized fields.

\begin{figure*}[!ht]
\centering
\includegraphics[width = 0.9\textwidth]{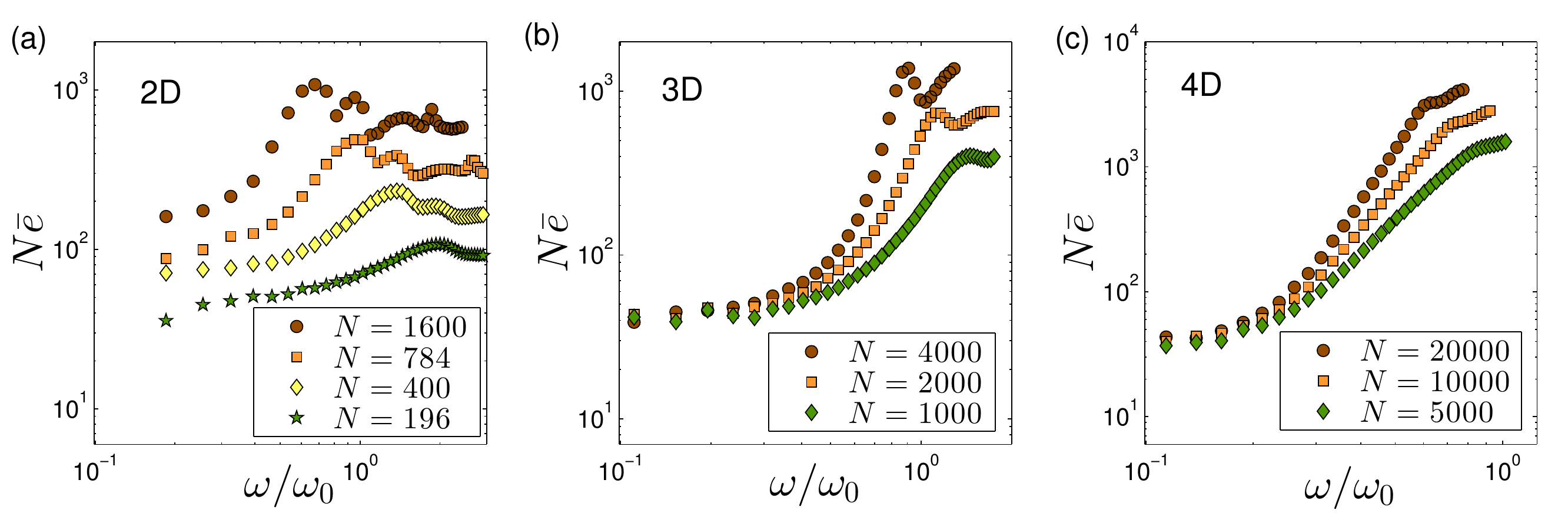}
\caption{\footnotesize Running averages $\bar{e}$ of the participation ratio $e$ (see text for definition) of low-frequency vibrational modes measured in (a) 2D, (b) 3D, and (c) 4D glasses, scaled by the system size $N$, binned over and plotted against frequency $\omega$. For visualization purposes the frequency axes are rescaled by the same scales $\omega_0$ as reported in the caption of Fig.~\ref{all_dimensions_dos}. The peaks of $\bar{e}$ correspond to the lowest frequency phonons \cite{phonon_widths}. }
\label{all_dimensions_participation}
\end{figure*}

Figure~\ref{all_dimensions_participation} presents the running averages $\bar{e}$ of the participation ratio $e$ of vibrational modes, representing here $\vec{z}$, binned over their frequencies $\omega$ and multiplied by the system size $N$. Panels (b) and (c) show data for 3D and 4D respectively, and demonstrate that the participation ratio of QLVM that populate the $\omega^4$ tails of the vDOS follow $e\!\sim\!N^{-1}$, while the 2D data displayed in panel (a) clearly show $e\!\nsim\!N^{-1}$. Building again on the observation that QLVM feature a $r^{-(\dbar-1)}$ spatial decay beyond a scale $\xi$, we estimate to leading order $e\!\sim\! N^{-1}$ for $\dbar\!\ge\!3$, but $e\!\sim\!(\log N)^2/N$ in 2D, in qualitative consistence with our observations.
%%%%%%%%%%%%%%%%%%%%%%%%%%%%%%%%%%%%%%%%%%%%%%%%%%%%%%%
%%%%%%%%%%%%%%%%%%%%%%%%%%%%%%%%%%%%%%%%%%%%%%%%%%%%%%

\emph{The cutoff frequency $\omega_c$.---} In addition to firmly establishing the universality of the non-phononic $\omega^4$ vDOS, the study of QLVM in 4D reveals an important frequency scale $\omega_c$ above which the $\omega^4$ scaling breaks down, that is difficult to observe cleanly in lower dimensions. As opposed to observations in 2D and 3D, where the lowest frequency phonons overlap the $\omega^4$ scaling regime, the breaking down of the universal $\omega^4$ scaling in 4D is {\em not} the result of the intrusion of phonons, but instead is seen as an intrinsic, $N$-independent, \emph{emergent} property of the vDOS.

\begin{figure}[!ht]
\centering
\includegraphics[width = 0.48\textwidth]{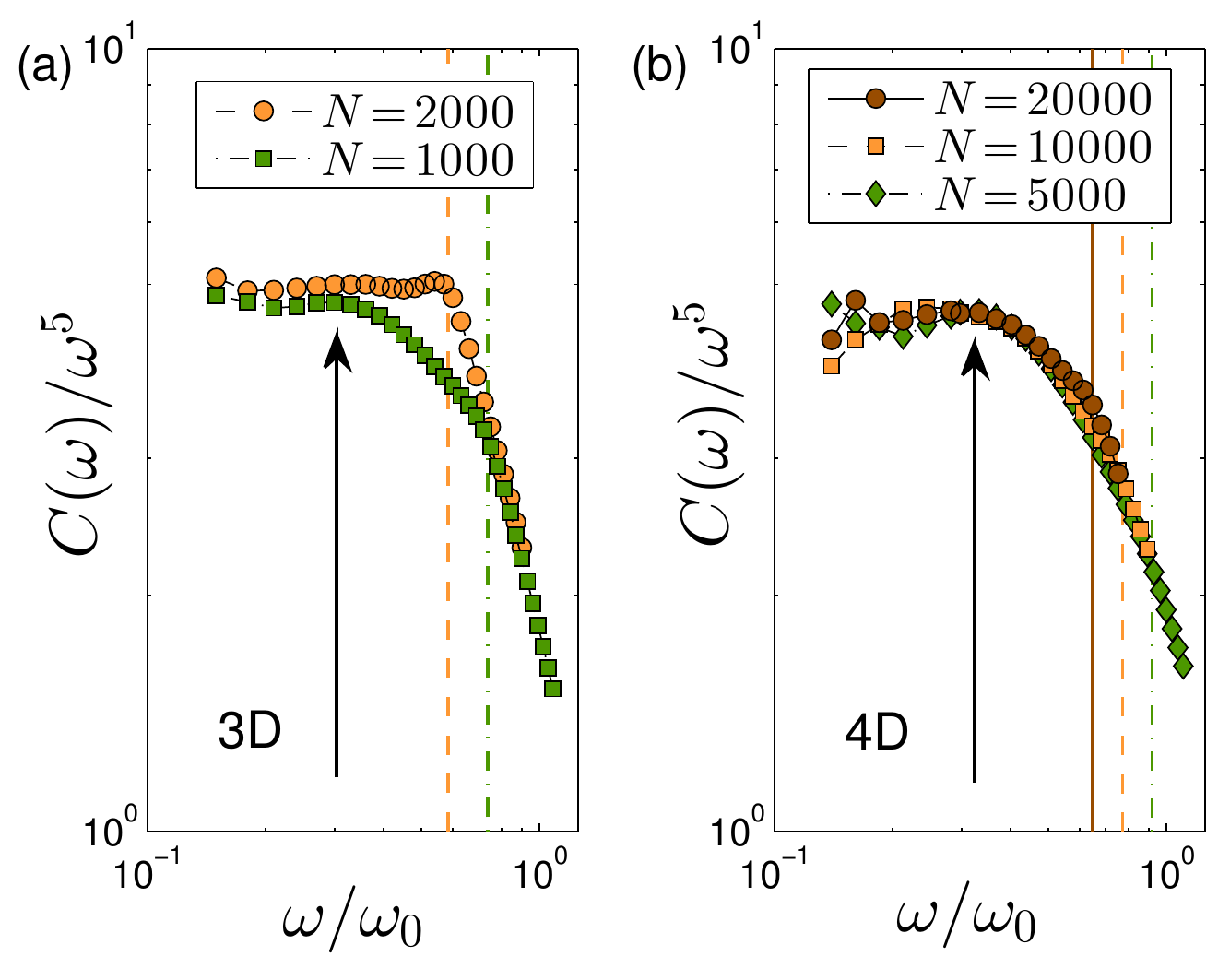}
\caption{\footnotesize Cumulative distributions $C(\omega)\!\equiv\!\int_0^\omega\! D(\omega')d\omega'$, divided by $\omega^5$ for (a) 3D systems and (b) 4D systems. The frequency axes are rescaled by $\omega_0\!=\!3.3$ (2.8) in 3D (4D) for visualization purposes. The vertical lines indicate the estimated position of the first phonon band, i.e.~they follow $2\pi \sqrt{G/\rho}/L$ with $G$ the athermal shear modulus. The cutoff frequency scale $\omega_c$ is marked by the vertical arrows, at the frequency at which $C(\omega)/\omega^5$ dips downwards from the low-frequency plateau, the latter reflecting the $\omega^4$ scaling of $D(\omega)$.}
\label{omega_1_fig}
\end{figure}

The emergence of $\omega_c$ is demonstrated in detail in Fig.~\ref{omega_1_fig}, where we plot the cumulative distributions $C(\omega)\!\equiv\!\int_0^\omega\! D(\omega')d\omega'$ rescaled by $\omega^5$, see caption for further details. Figure~\ref{omega_1_fig}b presents data calculated for 4D glasses, and clearly shows the emergence of the cutoff frequency $\omega_c$  --- the frequency above which the plateau of $C(\omega)/\omega^5$ ends, marked by the vertical arrow. Our 4D data indicate that $\omega_c$ is independent of system size. In the 3D data displayed in Fig.~\ref{omega_1_fig}a, the cutoff frequency $\omega_c$ can be only seen for $N\!=\!1000$; for $N\!=\!2000$ the first phonon band marked by the dashed horizontal line comes too close to $\omega_c$ and obstructs its visibility. We find $\omega_c/\omega_D\!\approx\!0.05$ for both 3D and 4D.

%%%%%%%%%%%%%%%%%%%%%%%%%%%%%%%%%%%%%%%%%%%%%%%%%%%%%%%
%%%%%%%%%%%%%%%%%%%%%%%%%%%%%%%%%%%%%%%%%%%%%%%%%%%%%%

\emph{Summary and discussion.---} In this Letter we showed that the non-phononic vDOS in structural glasses follows an $\omega^4$ law in 2D, 3D and 4D, firmly establishing its universality. We further demonstrated that the degree of localization of the modes that populate the $\omega^4$ tails of the vDOS is weak in 2D, giving rise to a pronounced system-size dependence of the prefactors of $\omega^4$ scaling of the vDOS, and to a scaling $e\!\nsim\! N^{-1}$ of the participation ratio $e$ of low-frequency quasilocalized vibrational modes, whereas $e\!\sim\! N^{-1}$ is found in 3D and 4D. Finally, we showed that by employing simulations in 4D we are able to cleanly identify a frequency scale $\omega_c$ above which the $\omega^4$ scaling of the vDOS breaks down.

% \emph{Summary and discussion.---} In this Letter we showed that the non-phononic vDOS in structural glasses follows an $\omega^4$ law in 2D, 3D and 4D, firmly establishing its universality. These results stand in contrast to mean-field approaches~\cite{eric_boson_peak_emt,silvio} that predict a dimension-independent non-phononic vDOS that follows $D(\omega)\!\sim\!\omega^2$. It remains to be seen whether for $\dbar\!>\!4$ deviations from the $\omega^4$ law are observed. In addition, our results
% rule out the scenario proposed in~\cite{Schirmacher_prl_2007} that the non-phononic vDOS follows an $\omega^{\dbar+1}$ law.

Our estimate $\omega_c/\omega_D\!\approx\!0.05$ may have interesting implications; if we take $T_D\!\approx\!200$ K as a rough estimate of the Debye temperature of many structural glasses, then $\omega_c$ translates into a temperature scale of $10$ K. This temperature scale appears to coincide with the temperature at which the specific heat of many structural glasses, when normalized by $T^3$ (Debye model's prediction), exhibits a maximum~\cite{Buchenau_prl_1993,ramos_2004}. It remains to be seen whether the similar magnitude of these two temperatures is deep or a mere coincidence. The robustness of $\omega_c$ should also be tested in the nonlinear excitations framework put forward in~\cite{plastic_modes_prerc, SciPost2016}.

The observed persistence of the $\omega^4$ law across spatial dimensions rules out the scenario proposed in~\cite{Schirmacher_prl_2007} that the non-phononic vDOS follows $D(\omega)\!\sim\!\omega^{\dbar+1}$, and stands in contrast to mean-field approaches~\cite{eric_boson_peak_emt,silvio} that predict a dimension-independent non-phononic vDOS that follows $D(\omega)\!\sim\!\omega^2$. It remains to be seen whether for $\dbar\!>\!4$ deviations from the $\omega^4$ law are observed. Finally, we note that recent computational advances support that the $\omega^4$ law persists in deeply annealed glasses \cite{LB_modes_2018}, albeit with a strongly suppressed prefactor $\omega_g^{-5}$. This observation is consistent with experiments on vapor-deposited glasses \cite{ediger_2017} suggesting the depletion of tunneling two-level systems, which are believed to be intimately related to quasilocalized excitations \cite{soft_potential_model_1991,marco2015prl}.

\textit{Acknowledgments.--} E.~L.~acknowledges support from the Netherlands Organisation for Scientific Research (NWO) (Vidi grant no.~680-47-554/3259). E.~B.~acknowledges support from the Minerva Foundation with funding from the Federal German Ministry for Education and Research, the William Z.~and Eda Bess Novick Young Scientist Fund and the Harold Perlman Family.

%\bibliography{references_lerner}


\end{document}